\documentclass[]{spie}  

\usepackage{amsmath,amsfonts,amssymb}
\usepackage{graphicx}
\usepackage[colorlinks=true,
citecolor=black,
linkcolor=black,
urlcolor=blue]{hyperref}
\usepackage{booktabs}

\usepackage{algorithm}
\usepackage{algpseudocode}
\algrenewcommand\textproc{}

\graphicspath{{figures/plots/}{figures/drawings/}}  

\usepackage{subcaption}	

\usepackage[mode=text]{siunitx}	
\DeclareSIUnit\noop{\relax}	
\DeclareSIUnit{\muas}{\unit{\micro\noop}as} 
\DeclareSIUnit{\mas}{\unit{\milli\noop}as}	
\DeclareSIUnit{\waves}{waves}
\usepackage[nolist,nohyperlinks]{acronym}
\usepackage{etoolbox}
\makeatletter
\newif\if@in@acrolist
\AtBeginEnvironment{acronym}{\@in@acrolisttrue}
\newrobustcmd{\LU}[2]{\if@in@acrolist#1\else#2\fi}  
\makeatother

\newacroindefinite{RMS}{an}{a}

\usepackage{tikz}
\usepackage{pgfplots} 

\definecolor{LGS}{RGB}  {255, 102,   0}
\definecolor{NGS}{RGB}  {0,    61, 230}
\definecolor{Mult}{RGB} {227,  31, 255}
\definecolor{GPRoffl}{RGB} {10,   150,   10}
\definecolor{GPRonl}{RGB} {0,   75,   0}
\definecolor{Butter}{RGB}{180,  0,   0}
\definecolor{Ref}{RGB}  {96,   96,  96}

\title{Simulation of Wavefront-based Disturbance Observers for Large Telescopes}

\author[a,b]{Pascal Jaufmann}
\author[a]{Aaron Buck}
\author[a]{Marco Zaiser}
\author[b]{J\"org-Uwe Pott}
\author[a]{Oliver Sawodny}
\affil[a]{Institute for System Dynamics, University of Stuttgart, Waldburgstr. 17/19, 70563 Stuttgart, Germany}
\affil[b]{Max Planck Institute for Astronomy, K\"onigsstuhl 17, 69117 Heidelberg, Germany}

\authorinfo{Further author information (Send correspondence to P.J.):\\
	Pascal Jaufmann, E-mail: pascal.jaufmann@isys.uni-stuttgart.de\\
	Jörg-Uwe Pott, E-mail: jpott@mpia.de
	}

\begin{document} 

\begin{acronym}[spaceeeeeeeeee] 
	\acro{ESO}	{European Southern Observatory}
	\acro{ELT}	{Extremely Large Telescope}
	\acro{VLT}	{Very Large Telescope}
	\acro{AOF}	{Adaptive Optics Facility}
	
	\acro{M1}	{\LU{P}{p}rimary \LU{M}{m}irror}
	\acro{M2}	{\LU{S}{s}econdary \LU{M}{m}irror}
	\acro{M3}	{\LU{T}{t}ertiary \LU{M}{m}irror}
	
	\acro{MICADO}	{Multi-AO Imaging Camera for Deep Observations}
	\acro{MORFEO}	{Multi-conjugate adaptive Optics Relay For ELT Observations}
	\acro{COMPASS}  {\LU{COM}{com}puting \LU{P}{p}latform for \LU{A}{a}daptive optic\LU{S}{s} \LU{s}{s}ystem}
	
	\acro{AO}	{\LU{A}{a}daptive \LU{O}{o}ptics}
	\acro{SCAO}	{\LU{S}{s}ingle-\LU{C}{c}onjugate \LU{A}{a}daptive \LU{O}{o}ptics}
	\acro{MCAO}	{\LU{M}{m}ulti-\LU{C}{c}onjugate \LU{A}{a}daptive \LU{O}{o}ptics}
	
	\acro{WFS}	{\LU{W}{w}ave\LU{F}{f}ront \LU{S}{s}ensor}
	\acro{WFE}	{\LU{W}{w}ave\LU{F}{f}ront \LU{E}{e}rror}
	\acro{DM}	{\LU{D}{d}eformable \LU{M}{m}irror}
	
	\acro{LGS}	{\LU{L}{l}aser \LU{G}{g}uide \LU{S}{s}tar}
	\acro{NGS}	{\LU{N}{n}atural \LU{G}{g}uide \LU{S}{s}tar}
	
	\acro{PSF}	{\LU{P}{p}oint \LU{S}{s}pread \LU{F}{f}unction}
	\acro{RMS}	{\LU{R}{r}oot \LU{M}{m}ean \LU{S}{s}quare}
	\acro{SVD}	{\LU{S}{s}ingular \LU{V}{v}alue \LU{D}{d}ecomposition}
	\acro{FoV}	{\LU{F}{f}ield of \LU{V}{v}iew}
	\acro{PtV}	{\LU{P}{p}eak-to-\LU{V}{v}alley}
	
	\acro{KF}	{Kalman \LU{F}{f}ilter}
	\acro{LQG}	{\LU{L}{l}inear \LU{Q}{q}uadratic \LU{G}{g}aussian \LU{C}{c}ontrol}
	
	\acro{PSD}	{\LU{P}{p}ower \LU{S}{s}pectral \LU{D}{d}ensity}
	\acro{GPR}	{Gaussian \LU{P}{p}rocess \LU{R}{r}egression}
	\acro{SE}	{\LU{S}{s}quared \LU{E}{e}xponential}
	\acro{AGWN} {\LU{A}{a}dditive Gaussian \LU{W}{w}hite \LU{N}{n}oise}
	\acro{iDFT}	{inverse \LU{D}{d}iscrete Fourier \LU{T}{t}ransformation}
	
	\acro{AR2}{\LU{A}{a}uto-\LU{R}{r}egressive process of order two}
	
	
	
\end{acronym}
	
\maketitle

\begin{abstract}
The performance of future observatories such as the \acl{ELT} is mainly limited by atmospheric turbulence and structural vibrations of the optical assembly.
To further enhance the mitigation performance of \acl{AO}, real-time information about the disturbances acting on the control loop is needed.
Current systems therefore employ a combination of \acl{WFS}- and accelerometer-based filters.
In this work, methods using only data from natural- and laser guide star (\acs{NGS}, \acs{LGS}) measurements are presented, as telescopes like the \acl{VLT} already have multiple fast and high-resolution \aclp{WFS} installed.
This approach also avoids the costly installation and operation of additional accelerometers on the optical elements.

We introduce two innovative disturbance observer schemes to sense both turbulence and vibration information.
A multi-rate estimator for atmospheric influences is based on Kalman filter theory and can incorporate \acs{NGS} and \acs{LGS} signals at different loop rates.
The estimator for structural perturbations uses \acl{GPR} and can be implemented in an offline and online configuration.
We validate the filter designs with data from a realistic end-to-end \acl{AO} model with randomly generated turbulence and vibrations.
The simulation is fed with on-sky data from the \acl{AOF} of the \acl{VLT}.
The presented disturbance observer schemes demonstrate promising results and may be considered as potential alternatives or extensions to existing techniques such as linear-quadratic controllers with Kalman filtering (\acs{LQG}).
\end{abstract}


\section{Introduction} \label{sec:Introduction}
The wavefronts of light from celestial objects are distorted by various influences on their path to the telescope. The primary sources of these disturbances is turbulence across different atmospheric layers and structural vibrations of the telescope, induced by wind and other mechanical forces. Adaptive optics (\acs{AO})\acused{AO} is the technology employed in modern telescopes to correct these perturbations. \ac{AO} systems utilize \acp{WFS} to measure the wavefront of incoming light from astronomical sources and employ \acp{DM} to correct these aberrations \cite{Hardy1998}.

\begin{figure}%
\centering
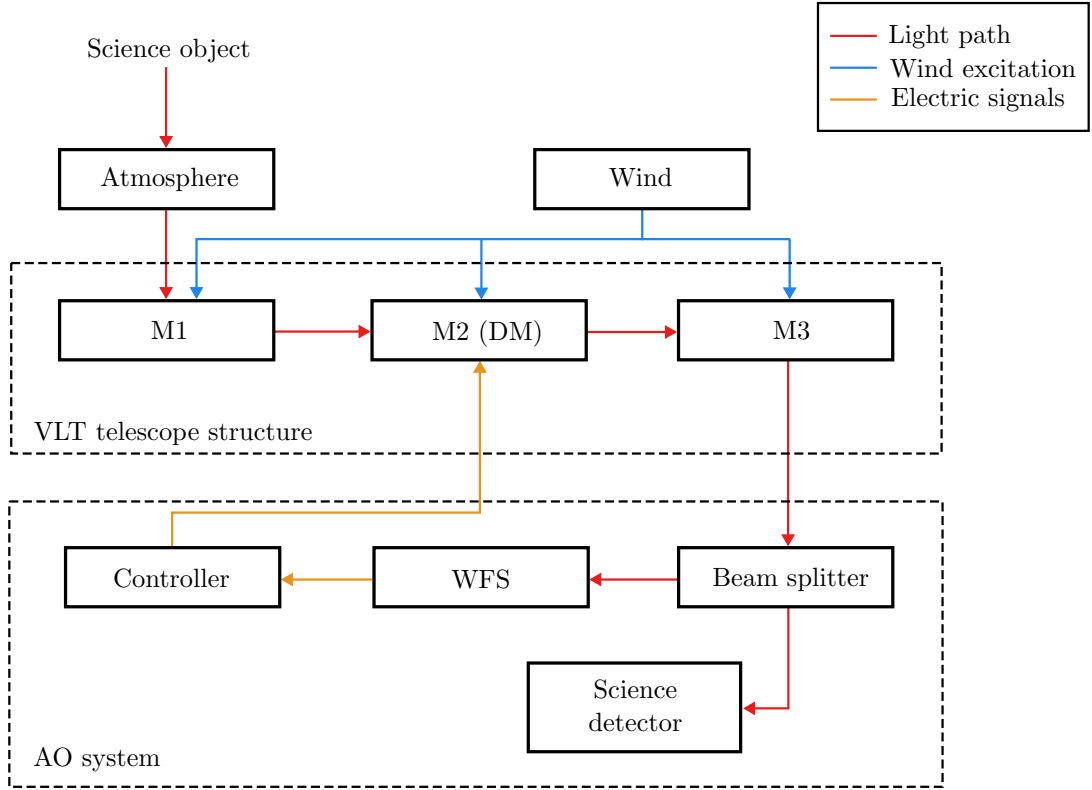
\caption{Overview of the light and signal paths through the \ac{VLT} UT4 (\emph{Yepun}) telescope structure.} 
\label{fig:VLT}
\end{figure}

Since disturbances caused by structural vibrations and turbulence are not directly measurable, methods to estimate these effects are required. In the configuration of the \ac{ESO} \ac{VLT}, light is initially captured by the \ac{M1}, subsequently reflected by a \ac{M2}, which functions as the \ac{DM}, and is finally directed by a \ac{M3} towards the scientific instruments. The \ac{M2} is mounted above \ac{M1} using lightweight structural supports, which are the primary contributors to the errors associated with structural vibrations. The way in which light travels through the \ac{VLT} mirrors and where the different errors are induced is illustrated in Figure~\ref{fig:VLT}. 
In this work, we compare the performance of a \ac{KF} and a Gaussian process based on \ac{LGS} wavefront data for estimating structural vibrations.
Additionally, we assess atmospheric turbulence also using a Kalman filter based solely on \ac{NGS} or \ac{LGS} data and a multi-rate observer that fuses \ac{NGS} and \ac{LGS} measurements.

To measure the distortion of wavefronts, light from a nearby \ac{NGS} is analyzed. To correct the high-frequent changes in the atmosphere, the \ac{AO} system needs to be operated with a frequency of 500-1000 Hz \cite{Rosensteiner2013}. Often, within the isoplanatic patch of the scientific target, a sufficiently bright \ac{NGS} to run the system on a high enough bandwidth cannot be found, necessitating the generation of \ac{LGS} to enhance the sky coverage of telescopes \cite{Hardy1998}. However, as the desired atmospheric windows are better covered by \ac{NGS}, due to \ac{LGS} issues like the cone effect \cite{Rosensteiner2013} and uplink turbulence \cite{AndrewReeves2016} and atmospheric tip-tilt indeterminacy \cite{AndrewReeves2016, Rosensteiner2013}, \ac{NGS} measurements must still be used in addition to \ac{LGS} measurements \cite{Hardy1998}. In new generations of \ac{AO} systems, known as \ac{MCAO}, multiple \ac{LGS} \ac{WFS} and potentially multiple \ac{DM} are employed to minimize the issues associated with \ac{LGS} and to increase sky coverage and \ac{FoV} \cite{Rosensteiner2013, Petit2009}. In the \ac{PSD} of \ac{WFS} data from large telescopes, significant peaks are observable at high frequencies in tip-tilt modes. Since disturbances due to atmospheric turbulence in tip-tilt are primarily in the low frequency range, these peaks can be identified with the natural frequencies of the telescopes \cite{Hardy1998, Keck2015}. In this simulation based on data of the \ac{VLT} UT4 (\emph{Yepun}) telescope, this occurs at 47,4969 Hz. As errors induced by structural vibrations grow with the size of telescopes, increasingly precise methods for estimating these vibrations are required \cite{Glueck2017}. Current correction methods for structural vibrations primarily utilize \ac{LQG} \cite{Correia2012} based on measurements from wavefront sensors \cite{Petit2008}, accelerometers mounted on telescope mirrors \cite{Keck2015}, or a combination of both \cite{Glueck2020}.

We present different observer designs for the estimation of atmospheric turbulence and structural vibrations.
Current research is investigating multi-rate methods that integrate data from wavefront sensors and accelerometers \cite{Glueck2020} for estimating mirror motions.
This work develops a similar multi-rate filter that merges measurements from \ac{NGS} and \ac{LGS} \acp{WFS}, aiming to enhance turbulence estimation and benchmarking it against more traditional Kalman filter designs.
Afterwards, we propose a new vibration estimation technique based on \ac{GPR} and compare it to observers based on a Kalman filter, as employed in \ac{LQG}.

The remainder of this paper is organized as follows: Section~\ref{sec:simulation} develops the \ac{AO} model for the baseline system and derives the equations governing the Kalman filter, the \ac{GPR}, and the multi-rate observer.
In Section~\ref{sec:turbulence}, the paper reports on the application of the Kalman filter and the multi-rate observer for estimating atmospheric turbulence, followed by an analysis of these findings.
Section~\ref{sec:vibrations} presents the simulation outcomes for the Kalman filter and the \ac{GPR} focused on estimating structural vibrations and discusses the implications of these results.
Finally, Section~\ref{sec:conclusion} summarizes the key discoveries and concludes the study.

\section{Simulation setup} \label{sec:simulation}
This section describes the setup of our \ac{AO} model, which utilizes data from a \acs{COMPASS} (\acl{COMPASS})\acused{COMPASS} simulation \cite{Ferreira2018} informed by observational data from the \ac{ESO} \ac{VLT}. The simulation employs two Shack-Hartmann \ac{WFS} specifically for capturing light from the \ac{LGS}, in addition to an auxiliary Shack-Hartmann \ac{WFS} for acquiring measurements from the \ac{NGS}. These measurements are utilized to generate the control signal~$u$. Corrections to optical disturbances are effected by a singular \ac{DM}, which typically comprises reflective plates that are dynamically deformed by actuators located beneath its surface.
\subsection{AO model}
\begin{figure}
	\centering
	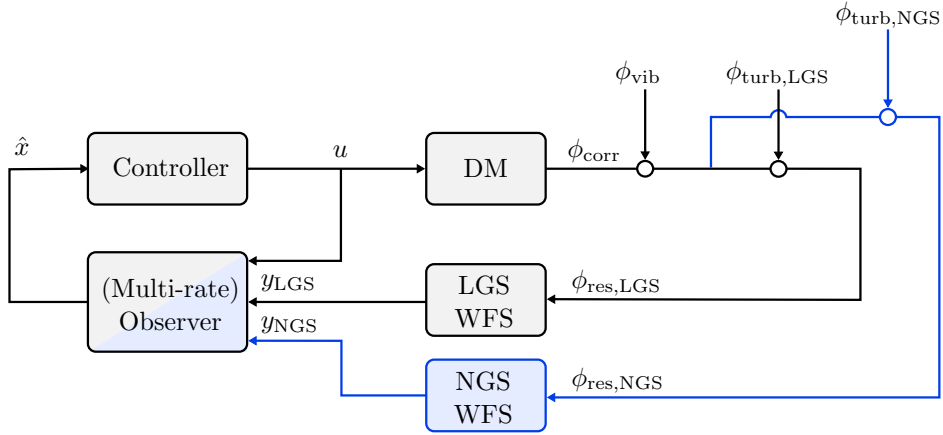%
	\caption{Typical \ac{AO}-structure with an observer.}
	\label{fig:Blockdiagram}
\end{figure}

An overview of an \ac{AO}-structure is visualized in Figure \ref{fig:Blockdiagram}. It includes structural ($\phi_\text{vib})$ and atmospheric ($\phi_\text{turb})$ disturbances as well as an observer for estimating all relevant control states $\boldsymbol{x}$. The blue color indicates the additional signal paths when a multi-rate observer is used.
\subsubsection*{Atmospheric turbulence}
The simulation adopts an atmospheric model based on Kolmogorov's stochastic description, which posits that solar energy input is systematically dissipated into increasingly finer structures, thus instigating energy cascades and subsequent turbulence \cite{Hardy1998}. Turbulence over brief intervals (less than one second) is characterized using Taylor's frozen flow hypothesis. According to this model, turbulence layers are assumed to be static and to traverse with constant velocities independent of each other \cite{Roddier1999}. The simulation is restricted to a single turbulence layer. The \ac{PSD} of this atmospheric model is approximated by a stochastic \ac{AR2} \cite{Meimon2010}:
\begin{subequations}\label{eq:turb_model}
\begin{align}
\phi_{\mathrm{turb}}[k+1]&=a_{\mathrm{turb},1}\phi_{\mathrm{turb}}[k]+a_{\mathrm{turb},2}\phi_{\mathrm{turb}}[k-1]+v_{\mathrm{turb}}[k]\label{eq:AR(2)Turb}, \\
a_{\mathrm{turb},1}&=2e^{-\xi \omega_0 T}cos(\omega_0\sqrt{1-\xi^2}T)\label{eq:AR1_Turb}, \\
a_{\mathrm{turb},2}&=-e^{-2\xi\omega_0 T}\label{eq:AR2_Turb}, \\	
\omega_0&=0.6\pi(n(i)+1)\frac{V_0}{D}. \label{eq:w_0}
\end{align}
\end{subequations}
The parameters assigned for the simulation are delineated in Table~\ref{tab:Turb_Parameter}.
\begin{table} 
	\centering
	\caption{Parameters of the turbulence model.}
	\begin{tabular}{lll}
		\toprule
		Parameter& Name & Value \\
		\midrule
		$\xi$ & Damping factor & $0.9$ \\
		$n(i)$ & Radial order of the Zernike mode i & $1$ for tip-tilt ($i=1$) \\
		$T$	& Exposure time & \qty{1}{ms} for \ac{LGS}, \qty{5}{ms} for \ac{NGS}\\
		$V_0$ & Wind velocity & \qty{15}{m/s}\\
		$D$ & Aperture diameter & \qty{8}{m}\\
		\bottomrule
	\end{tabular}
	\label{tab:Turb_Parameter}
\end{table}
\subsubsection*{Structural vibrations}
Disturbances resulting from structural vibrations are viewed as supplementary deformations of the wavefront, additional to the atmospheric disturbances, particularly in the tip-tilt Zernike modes. These vibrations are conceptualized as a mass-spring-damper system. The corresponding dynamic equations are represented again through an \ac{AR2} model \cite{Meimon2010}:
\begin{subequations}\label{eq:vib_model}
\begin{align}
\phi_{\mathrm{vib}}[k+1]&=a_{\mathrm{vib},1}\phi_{\mathrm{vib}}[k]+a_{\mathrm{vib},2}\phi_{\mathrm{vib}}[k-1]+v_{\mathrm{vib}}[k]\label{eq:AR(2)Vib}, \\
a_{\mathrm{vib},1}&=2e^{-d \omega_\text{f} T}cos(\omega_f\sqrt{1-\xi_\text{f}^2}T)\label{eq:AR1Vib}, \\
a_{\mathrm{vib},2}&=-e^{-2d\omega_\text{f} T}	\label{eq:AR2Vib}.
\end{align}
\end{subequations}
The selection of the parameters is detailed in Table~\ref{tab:Vib_Parameter}. The parameters~$d$ and~$\omega_\text{f}$ are derived from the \ac{PSD} of the \ac{NGS} \ac{WFS} data, see Figure~\ref{fig:VLT_NGS_tip_PSD}.

\begin{figure}
	\centering
	\input{figures/plots/VLT_NGS_tip_PSD.tikz}
	
	\caption[\ac{PSD} of \ac{VLT} \ac{NGS} \ac{WFS} data.]{\ac{PSD} of real \ac{VLT} \ac{NGS} \ac{WFS} data including some natural frequencies.
		\ac{VLT} \ac{NGS} \ac{WFS}  (\tikz[baseline=-0.5ex]\draw[NGS, thick] (0,0) -- (0.5,0);),
		dominant natural frequency at \qty{47.4969}{\Hz} (\protect\tikz[baseline=-0.5ex]\protect\draw[Ref, thick, dash pattern=on 2pt off 2pt] (0,0) -- (0.5,0);),
		additional frequencies for the Butterworth filter  (\tikz[baseline=-0.5ex]\draw[red, thick] (0,0) -- (0.5,0);).
	}
	\label{fig:VLT_NGS_tip_PSD}
\end{figure}

\begin{table} 
	\centering
	\caption{Parameters of the vibration model.}
	\begin{tabular}{lll}
		\toprule
		Parameter& Name & Value   \\
		\midrule
		$\omega_\text{f}$ & Natural frequency of the vibration & \qty{47.4969}{\Hz}\\
		$\xi_\text{f}$ & Damping factor & 0.03\\
		$T$	& Exposure time & \qty{1}{\ms} for \ac{LGS}, \qty{5}{\ms} for \ac{NGS}\\
		\bottomrule
	\end{tabular}
	\label{tab:Vib_Parameter}
\end{table}

\subsubsection*{\ac{DM} and \ac{WFS} dynamics}
The wavefront error $y$, measured by the Shack-Hartmann \ac{WFS}, is interpreted as a projection of the modal wavefront error into the sensor space, augmented by \ac{AGWN}~$w$ \cite{Glueck2017}
\begin{equation}
y[k]=D\phi_{\mathrm{res}}[k-1]+w[k].
\label{eq:Gemessene Wellenfront}
\end{equation}
The residual wavefront error~$\phi_{\mathrm{res}}$ comprises two components. The wavefront~$\phi_{\mathrm{dist}}$ includes disturbances from both atmospheric turbulence and structural vibrations.
The correction component~$\phi_{\mathrm{corr}}$ is attributable to the wavefront adjustments made by the \ac{DM}:
\begin{subequations}
\begin{align}
\phi_{\mathrm{res}}[k-1]&=\phi_{\mathrm{dist}}[k-1]+\phi_{\mathrm{corr}}[k-1]\label{eq:phi_res}, \\
\phi_{\mathrm{dist}}[k]&=\phi_{\mathrm{turb}}[k]+\phi_{\mathrm{vib}}[k]\label{eq:phi_dist}.
\end{align}
\end{subequations}
The internal position control of the \ac{DM} actuators is sufficiently faster than the wavefront control of the \ac{AO}-system. We thus assume, that the \ac{DM} is in a steady state and model it as a delay of one time step. The relation between the corrected modal wavefront~$\phi_{\mathrm{corr}}$ and the \ac{DM} input~$u$ is described with a projection~$N$:
\begin{equation}
\phi_{\mathrm{corr}}[k-1]=Nu[k-2].
\label{eq:DM Modale Wellenfront}
\end{equation}
In our simulation, both the data from the \ac{WFS} and the control signal $u$ are already represented in Zernike modes. Consequently, we set $D = I^{n\times n}$ and $N = I^{n\times n}$.
\subsubsection*{Overall model}
For each Zernike mode, the system dynamics are encapsulated within the state space vector 
\begin{equation}
\boldsymbol{x}[k+1]=
\begin{bmatrix}
    \phi_{\mathrm{turb}}[k] \\
    \phi_{\mathrm{turb}}[k-1] \\
    \phi_{\mathrm{vib}}[k] \\
    \phi_{\mathrm{vib}}[k-1] \\
    u[k-1] \\
    u[k-2]
\end{bmatrix}. \label{eq:Zustandsvektor}
\end{equation}
Zernike polynomials are not strictly stochastically independent, but the small correlation can be neglected in the lower Zernike modes~\cite{Hardy1998}. For this reason we can model tip and tilt separately with the same state matrices.
The resulting system will thus be described by the discrete state space model
\begin{subequations}\label{eq:model_x}
\begin{align}
\boldsymbol{x}[k+1]&=\boldsymbol{A}\boldsymbol{x}[k]+\boldsymbol{B}u[k]+v[k]\label{eq:x_dot}, \\
y[k]&=\boldsymbol{C}\boldsymbol{x}[k]+w[k]\label{eq:y},
\end{align}\\[-3em]
\begin{align}
&\boldsymbol{A}=
\begin{bmatrix}
	\boldsymbol{A}_{\mathrm{turb}}&0&0\\
	0&\boldsymbol{A}_{\mathrm{vib}}&0\\
	0&0&\boldsymbol{A}_u	
\end{bmatrix}\label{eq:A},\qquad
&&\boldsymbol{A}_{\mathrm{turb}}=
\begin{bmatrix}
	a_{\mathrm{turb},1}&a_{\mathrm{turb},2}\\
	1&0
\end{bmatrix},\quad
\boldsymbol{A}_{\mathrm{vib}}=
\begin{bmatrix}
	a_{\mathrm{vib},1}&a_{\mathrm{vib},2}\\
	1&0
\end{bmatrix},\quad
\boldsymbol{A}_{\mathrm{u}}=
\begin{bmatrix}
	0&0\\
	1&0
\end{bmatrix},\\
&\boldsymbol{B}=
\begin{bmatrix}
	\boldsymbol{B}_\mathrm{turb}\\\boldsymbol{B}_\mathrm{vib}\\\boldsymbol{B}_{u}
\end{bmatrix},\qquad
&&\boldsymbol{B}_\mathrm{turb}=\begin{bmatrix}0\\0\end{bmatrix},\quad
\boldsymbol{B}_\mathrm{vib}=\begin{bmatrix}0\\0\end{bmatrix},\quad
\boldsymbol{B}_\mathrm{u}=\begin{bmatrix}0\\1\end{bmatrix},\\
&\boldsymbol{C}=
\begin{bmatrix}
	\boldsymbol{C}_\mathrm{turb} & \boldsymbol{C}_\mathrm{vib} & \boldsymbol{C}_{u}
\end{bmatrix},\qquad
&&\boldsymbol{C}_\mathrm{turb}=\begin{bmatrix}0&1\end{bmatrix},\quad
\boldsymbol{C}_\mathrm{vib}=\begin{bmatrix}0&1\end{bmatrix},\quad
\boldsymbol{C}_{u}=\begin{bmatrix}0&1\end{bmatrix}. \label{eq:C}
\end{align}
\end{subequations}
The process noise $v[k]$ and measurement noise $w[k]$ represent \ac{AGWN}.

\subsection{Kalman filtering} \label{ssec:KF}
From the state space model described, we proceed to derive a Kalman filter.
The filter will operate on either \ac{LGS} or \ac{NGS} measurements.
For the specific task of estimating structural vibrations, we will utilize the \ac{LGS}-based Kalman filter due to the shorter exposure times of the  \ac{LGS} sensor, which allows for a greater bandwidth to follow the high-frequency vibrations.
Furthermore, \ac{AO} systems must be capable of estimating and correcting wavefront errors in real-time, which imposes conditions on the calculate time of the observers.
Since all system matrices are time-independent, we can use a steady-state Kalman filter design to minimize the computational demand.
The calculation of the Kalman gain $\boldsymbol{K}$ is detailed in equations~\eqref{eq:KF1} and can be done offline beforehand.
\begin{subequations}\label{eq:KF1}
\begin{align}
\boldsymbol{Q}&=E\{v[k]^Tv[k]\}, \qquad \boldsymbol{R}=E\{w[k]^Tw[k]\}\label{eq:Q und R}, \\
\boldsymbol{P}&=\boldsymbol{A}\boldsymbol{P}\boldsymbol{A}^T-(\boldsymbol{A}\boldsymbol{P}\boldsymbol{C}^T)(\boldsymbol{C}\boldsymbol{P}\boldsymbol{C}^T+\boldsymbol{R})^{-1}(\boldsymbol{C}\boldsymbol{P}\boldsymbol{A}^T)+\boldsymbol{Q} \label{eq:P}, \\
\boldsymbol{K}&=\boldsymbol{A}\boldsymbol{P}\boldsymbol{C}^T(\boldsymbol{C}\boldsymbol{P}\boldsymbol{C}^T+\boldsymbol{R})^{-1}. \label{eq:K}
\end{align}
\end{subequations}
The estimated states $\hat{\boldsymbol{x}}$ are then given by
\begin{subequations}\label{eq:KF2}
\begin{align}
\hat{\boldsymbol{x}}[k+1]&=\boldsymbol{A}\,\hat{\boldsymbol{x}}[k] + \boldsymbol{B}\,u[k] + \boldsymbol{K}(\boldsymbol{y}[k]-\boldsymbol{C}\hat{\boldsymbol{x}}[k]), \label{eq:x_dach} \\
\hat{y}[k]&=\boldsymbol{C}\,\hat{\boldsymbol{x}}[k]. \label{eq:y_dach}
\end{align}
\end{subequations}

\subsubsection{Multi-rate system}
A multi-rate observer based on \ac{LGS} data faces the challenge that in reality, \ac{LGS} cannot accurately determine tip and tilt modes of turbulence.
In \ac{COMPASS} however, there is no tip-tilt indetermination \cite{Ferreira2018}, allowing a multi-rate observer to be built, which can then be extended to higher modes where it may provide advantages in reality.
The multi-rate observer design adopts a sequential Kalman filter approach as proposed by Kulcsar \cite{Raynaud2008, Kulcsar2010}.
This methodology aligns with similar multi-rate strategies for estimating vibrations using a combination of accelerometer and \ac{WFS} data, as explored by Gl\"uck \cite{Glueck2020}.
The sequential implementation is possible as the exposure time of the \ac{NGS} \ac{WFS}, denoted as $T_{\text{NGS}}$, is an integer multiple of the \ac{LGS} \ac{WFS} exposure time, specifically $T_{\text{NGS}} = 5T_{\text{LGS}}$.
The goal of this approach is to design an observer that primarily follows the atmospheric conditions of the \ac{NGS}, while also estimating high-frequency disturbances that can only be measured by the faster \ac{LGS} \ac{WFS}.
We therefore prioritize the \ac{NGS} measurements due to their higher reliability, while also utilizing the \ac{LGS} measurements for capturing high-frequency components.
\begin{table} 
\caption{Overview of the \ac{WFS}.}
\label{fig:WFS}
\centering
\begin{tabular}{llll}
	\toprule
	\ac{WFS}& Type & Exposure time & Sampling rate   \\
	\midrule
	\ac{NGS} & Shack-Hartmann & 1 \unit{ms} & 200 \unit{Hz} \\
	\ac{LGS} & Shack-Hartmann & 5 \unit{ms} & 1000 \unit{Hz} \\
	\bottomrule
\end{tabular}
\end{table} 
This means that our multi-rate observer operates with the same bandwidth as the standard \ac{LGS} Kalman filter, incorporating \ac{NGS} measurements at every fifth time step. The Kalman gain $\boldsymbol{K}_{\mathrm{LGS}}$ for intervals featuring only \ac{LGS} measurements is computed using the equations \eqref{eq:KF1} detailed in section \ref{ssec:KF}. During time steps that include both \ac{LGS} and \ac{NGS} measurements, the Kalman gain $\boldsymbol{K}_{\mathrm{LGS}\wedge \mathrm{NGS}}$ is recalculated using the same foundational equations but with adjusted matrices for $\boldsymbol{C}$ and $\boldsymbol{R}$:
\begin{align}
\boldsymbol{R}&=
\begin{cases}
r_{\text{LGS}}, & kT_{\mathrm{LGS}} \text{ mod } T_{\mathrm{NGS}} \neq 0 \\[1em]
\begin{bmatrix}
	r_{\text{LGS}}&0\\
	0&r_{\text{NGS}}
\end{bmatrix},&\text{else}.
\end{cases}\label{eq:R_mult}
\end{align}
%
In the multi-rate context, we denote the vectors and matrices associated with intervals featuring only \ac{LGS} measurements with the subscript "\ac{LGS}", and those pertaining to time steps with both \ac{LGS} and \ac{NGS} measurements with the subscript "$\text{\ac{LGS}}\wedge\text{\ac{NGS}}$". We assign $r_{\text{\ac{NGS}}} < r_{\text{\ac{LGS}}}$, reflecting a lower variance for $r_{\text{\ac{NGS}}}$ based on the presumption that \ac{NGS} measurements provide a more accurate atmospheric representation. The estimated state is updated according to 
\begin{subequations}\label{eq:Multi-rate1}
\begin{align}
\hat{\boldsymbol{x}}_{\mathrm{pred}}[k+1]&=\boldsymbol{A}\hat{\boldsymbol{x}}[k]+\boldsymbol{B} u[k]\label{eq:x_pred_mult}, \\
\hat{\boldsymbol{x}}[k+1]&=
\begin{cases}
\hat{\boldsymbol{x}}_{\mathrm{pred}}[k+1]+\boldsymbol{K}_{\mathrm{LGS}}(\boldsymbol{y}_{\mathrm{LGS}}[k] - \boldsymbol{C}_{\mathrm{LGS}}\hat{\boldsymbol{x}}_{\mathrm{pred}}[k+1]), & kT_{\mathrm{LGS}} \text{ mod } T_{\mathrm{NGS}} \neq 0 \\[1em]
\hat{\boldsymbol{x}}_{\mathrm{pred}}[k+1] + \boldsymbol{K}_{\mathrm{LGS}\wedge \mathrm{NGS}}(\boldsymbol{y}_{\mathrm{LGS}\wedge \mathrm{NGS}}[k] - \boldsymbol{C}_{\mathrm{LGS}\wedge \mathrm{NGS}}\hat{\boldsymbol{x}}_{\mathrm{pred}}[k+1]), & \mathrm{else},
\end{cases}\label{eq:x_corr_mult}
\end{align}
\end{subequations}
with
\begin{align}
\boldsymbol{y}_{\mathrm{LGS}\wedge \mathrm{NGS}} = \begin{bmatrix}
	y_{\text{LGS}} \\
	y_{\text{NGS}}
\end{bmatrix}, \quad
\boldsymbol{C}_{\mathrm{LGS}\wedge \mathrm{NGS}} = \begin{bmatrix}
	0&1&0&1&0&1\\
	0&1&0&1&0&1
\end{bmatrix}.
\end{align}
\subsection{Gaussian-Process Regression filtering}
The \ac{GPR} is a non-parametric approach that operates without reliance on an explicit state space model\cite{Schoelkopf2018}.
The \ac{GPR} receives input data defined as $\boldsymbol{x}=\begin{bmatrix}u & y_{\text{LGS}}\end{bmatrix}$, and requires knowledge of the real vibration as a reference signal to facilitate the learning of the relationship between the input data and the disturbances.
The reference for real vibrations is obtained through a filtering process applied to the signal $\phi_{\text{dist}}$.
For this, a Butterworth filter is applied \cite{Butterworth1930}.
Based on the learning mechanism, the \ac{GPR} constructs a predictive model and attempts to estimate the vibrations using a model, which is formulated with a Gaussian kernel.
We therefore employ the \ac{SE} kernel\cite{Duvenaud2014}
\begin{equation}
k_{\text{SE}}(\boldsymbol{x}, \boldsymbol{x}') = \sigma_{\text{SE}}^2 \exp\left(-\frac{(\boldsymbol{x} - \boldsymbol{x}')^2}{2\,l^2}\right). \label{eq:Kernel}
\end{equation} 
This kernel evaluates the correlation between two data points $\boldsymbol{x}$ and $\boldsymbol{x}'$, where the parameter $l$ governs the sensitivity of the function to changes in the input data.
A smaller $l$ leads to a function that quickly responds to data variations, whereas a larger $l$ results in a smoother response function.
The variance of the output, denoted by $\sigma_{\text{SE}}^2$, acts as a scaling factor.
The \ac{GPR} can be implemented either as an offline algorithm, where training is conducted in advance using an existing dataset, or online, where the model is continuously updated with new data at each time step.

\subsubsection*{Butterworth filter}
As mentioned earlier, the \ac{GPR} needs the reference signal $\phi_{\text{vib,filt}}$ for the vibrations acting on the telescope structure.
Thus, a Butterworth filter is used to isolate the vibrations at specific frequencies from the signal~$\phi_{\text{dist}}$.
This is achieved by constructing multiple bandpass filters around the desired natural frequencies.
Based on the \ac{PSD} of the unfiltered \ac{NGS} \ac{WFS} data (see Figure~\ref{fig:VLT_NGS_tip_PSD}), we opt to filter out vibrations at six frequencies as listed in Table~\ref{tab:gpr_params}.

The filtering process is used both in the offline and online \ac{GPR}.
Consequently, $\phi_{\text{dist}}$ has to be approximated according to the respective setting.
For the offline \ac{GPR}, we use the open-loop \ac{LGS} \ac{WFS} measurement, i.e. $\phi_{\text{dist}}=y_{\text{\ac{LGS},OL}}$. This is reasonable, because open-loop sensor signals only contain disturbance information.
For the online \ac{GPR}, we approximate the open-loop signal with $\phi_{\text{dist}}=y_{\text{\ac{LGS},CL}} - u$.
We specifically exclude \ac{NGS} data in the vibration filtering, as the faster \ac{LGS} \ac{WFS} captures high-frequency vibration more accurately compared to the slower \ac{NGS} \ac{WFS}.
The estimation of real vibrations by filtering~$\phi_{\text{dist}}$ is likely to be better when open-loop data is used, as the \ac{DM} does not mitigate the disturbances seen by the \ac{WFS}.
Additionally, the absence of high-frequency movements from the \ac{DM} ensures less disturbance in the signal.
The filtered vibrations~$\phi_{\text{vib,filt}}$ using both open and closed-loop data are depicted in Figure \ref{fig:butter1}.
As predicted, the filtering using open-loop data is better with \iac{RMS} error of \num{7.66e-3} compared to \iac{RMS} error of \num{13.4e-3} when using closed-loop \ac{WFS} data.
\begin{figure}
	\centering
	\begin{subfigure}[b]{\textwidth}
		\centering
		\input{figures/plots/Butter_ol.tikz}
		\input{figures/plots/Butter_ol_zoom.tikz}
		\caption{Filtering with open-loop data, bottom plot zoomed.}
	\end{subfigure}
	\begin{subfigure}[b]{\textwidth}
		\centering
		\input{figures/plots/Butter_cl.tikz}
		\input{figures/plots/Butter_cl_zoom.tikz}
		\caption{Filtering with closed-loop data, bottom plot zoomed.}
	\end{subfigure}

	\caption[Butterworth filtering of $\phi_\text{dist}$ as vibration reference.] {
		Butterworth filtering of $\phi_\text{dist}$ as vibration reference (only tip mode shown).
		Filtered vibration $\phi_{\text{vib,filt}}$ (\protect\tikz[baseline=-0.5ex]\protect\draw[Butter, thick] (0,0) -- (0.5,0);)
		and real vibration $\phi_\text{vib}$(\protect\tikz[baseline=-0.5ex]\protect\draw[Ref, thick, dash pattern=on 2pt off 2pt] (0,0) -- (0.5,0);).
	}
	\label{fig:butter1}
\end{figure}

\subsubsection{Offline \ac{GPR}}
The offline configuration of the \ac{GPR} involves a single training phase prior to deployment, where the model is constructed.
This pre-trained model is subsequently used to estimate the vibrations.
The reference vibration~$\phi_{\text{vib,filt}}$ is obtained by Butterworth filtering the open-loop signal~$\phi_{\text{dist}}=y_{\text{\ac{LGS},OL}}$.
Initially, the data point~$\boldsymbol{x}$ must be defined, which is achieved using equation~\eqref{eq:x Definition}, incorporating $d$-amount of past control signals~$u$ and measurements~$y_\text{\ac{LGS}}$.
\begin{equation}
\boldsymbol{x} = \begin{bmatrix}
u[k - d + 1] & y_\text{\ac{LGS}}[k - d + 1] \\
\vdots & \vdots \\
u[k] & y_\text{\ac{LGS}}[k]
\end{bmatrix}
\label{eq:x Definition}
\end{equation}
The training process involves the derivation of the covariance matrix~$\boldsymbol{K}$ using the \ac{SE}-kernel.
To construct~$\boldsymbol{K}$, the kernel function operates on a three-dimensional matrix containing the training data~$\boldsymbol{x}_m$.
The cat() function is used to concatenate these matrices along the third dimension.
The offline \ac{GPR} training is depicted in Algorithm~\ref{alg:offlineGPRtrain}.
\begin{algorithm}
\caption{Offline \ac{GPR}: Training}
\begin{algorithmic}[1] 
\For{$k = d$ \textbf{to} $\text{train size} + d - 1$}
    \State $\boldsymbol{x}_m \gets \text{cat}(3, \boldsymbol{x}_m, \boldsymbol{x}[k - d + 1 : k])$
\EndFor
\State $\mathbf{K} \gets k_\text{SE}(\boldsymbol{x}_m, \boldsymbol{x}_m, \sigma_{\text{SE}}, l)$
\end{algorithmic}\label{alg:offlineGPRtrain}
\end{algorithm}
Upon completion of the training phase, the model can be employed to estimate structural vibrations using a new, unknown dataset as the test set.
Every time step, a new matrix~$\boldsymbol{x}_{m+1}$ containing the current \ac{WFS} measurement~$y_\text{\ac{LGS}}[k]$ and control input~$u[k]$ is utilized to compute the corresponding kernel vector~$\boldsymbol{k}$.
To make the next vibration prediction~$\hat{\phi}_{\text{vib}}[k]$, the covariance matrix~$\boldsymbol{K}$ is adjusted by the noise level~$\sigma_{\text{agwn}}$ and subsequently inverted.
Finally, the transpose of the kernel vector is multiplied by the inverse of the adjusted covariance matrix and the filtered reference signal~$\phi_{\text{vib,filt}}$.
This process is outlined in Algorithm~\ref{alg:offlineGPRtest} and the parameters used are displayed Table~\ref{tab:gpr_params}.
\begin{algorithm}
\caption{Offline \ac{GPR}}
\begin{algorithmic}[1] 
\For{$k = d$ \textbf{to} test size}
    \State $\boldsymbol{x}_{m+1} \gets \boldsymbol{x}[k - d + 1 : k]$
    \State $\boldsymbol{k} \gets k_\text{SE}(\boldsymbol{x}_m, \boldsymbol{x}_{m+1})$
    \State $\hat{\phi}_{\text{vib}}[k] \gets \boldsymbol{k}^T \cdot (\boldsymbol{K} + \sigma^2_{\text{awgn}} \cdot \boldsymbol{I})^{-1} \cdot \phi_{\text{vib,filt}}[d : \text{train size} + d]$
\EndFor
\end{algorithmic}\label{alg:offlineGPRtest}
\end{algorithm}

\subsubsection{Online \ac{GPR}}
In online \ac{GPR}, the model is continuously updated to incorporate new information at each time step.
Unlike its offline counterpart, which learns the model in advance, online \ac{GPR} is trained iteratively at every time step, utilizing data from $m$-number of recent and historical data points.
This method involves not only determining the kernel vector $\boldsymbol{k}$ but also recalculating the covariance matrix $\boldsymbol{K}$ in each iteration.
This iterative updating significantly increases computational time.
However, the overall approach remains analogous to offline \ac{GPR}, see Algorithm~\ref{alg:onlineGPR}.
The parameters used are listed in Table~\ref{tab:gpr_params}.
\begin{algorithm}[H]
\caption{Online \ac{GPR}}
\begin{algorithmic}[1] 
\For{$k = d + m$ \textbf{to} testsize}
    \For{$i = m$ \textbf{to} $1$}
        \State $\boldsymbol{x}_m \gets \text{cat}(3, \boldsymbol{x}_m, \boldsymbol{x}[k - i + d + 1 : k - i])$
    \EndFor
    \State $\boldsymbol{x}_{m+1} \gets \boldsymbol{x}[k - d + 1 : k]$
    \State $\boldsymbol{K} \gets k_\text{SE}(\boldsymbol{x}_m, \boldsymbol{x}_m)$
    \State $\boldsymbol{k} \gets k_\text{SE}(\boldsymbol{x}_m, \boldsymbol{x}_{m+1})$
    \State $\hat{\phi}_{\text{vib}}[k] \gets \boldsymbol{k}^T \cdot (\boldsymbol{K} + \sigma^2_{\text{agwn}} \cdot \boldsymbol{I})^{-1} \cdot \phi_{\text{vib,filt}}[k - m : k - 1]$
\EndFor
\end{algorithmic}\label{alg:onlineGPR}
\end{algorithm}

\begin{table}[H]
	\centering
	\caption{Parameters for the offline and online \ac{GPR}}
	\label{tab:gpr_params}
	\begin{tabular}{@{}lcc@{}}
		\toprule
		& \textbf{Offline \ac{GPR}} &  \textbf{Online \ac{GPR}}\\ 
		\midrule
		\textbf{Butterworth filter} & \hspace{3cm} & \hspace{3cm} \\
		Frequencies [Hz] & \multicolumn{2}{c} {17.1, 24.3, 34.1, 47.7, 54.3, 61.5} \\
		Bandwidth & 3 & 2\\
		Filter order & 1 & 1 \\
		\midrule
		\textbf{SE-kernel} & &\\
		Output variance $\sigma_{\text{SE}}$ & 10 & 30 \\
		Scaling factor $l$ & 300 & 300 \\
		\midrule
		\textbf{\ac{GPR}} & & \\
		Training size $m$ & 1000  & 511 \\
		Depth $d$ & 407 & 170 \\
		\ac{AGWN} $\sigma_{\text{agwn}}$ & 0.13  & 0.1 \\
		\bottomrule
	\end{tabular}
\end{table}

\subsection{Modeling vibrations}
To simulate the vibrations acting on the \ac{AO} system, we introduce an artificial disturbance $\phi_\text{vib}$.
This signal is generated based on the known \ac{PSD} $S(\omega)$ of the measured \ac{NGS} \ac{WFS} data at the \ac{VLT} (see Figure~\ref{fig:VLT_NGS_tip_PSD}).
It includes a dominant peak at \qty{47.4969}{\Hz} and several smaller peaks between \qty{20}{\Hz} and \qty{75}{\Hz}.
Frequencies above the cut-off frequency $\omega_\text{c}=\qty{80}{\Hz}$ are set to zero, since we assume, that these higher frequencies do not correspond to structural vibrations in the recorded \ac{PSD}.
Using the \ac{iDFT} with a random phase between $0$ and $2\,\pi$, we can calculate a time signal $\phi_\text{vib}$, shown in Algorithm~\ref{alg:iFFT}.
In our \ac{COMPASS} simulation, we then introduce the random vibration disturbance $\phi_\text{vib}$ similar to the atmospheric turbulence to the \ac{AO} loop, c.f. Figure~\ref{fig:Blockdiagram}.
\begin{algorithm}[h]
	\caption{\Acl{iDFT}}
	\begin{algorithmic}[1] 
		\Function{$\phi_\text{vib}=\ac{iDFT}$}{$S(\omega),\,\omega_\text{c}$}
		\State $S_\text{amp} \gets [S(1:\omega_\text{c}),\, 0\dots0 ]$
		\State $S_\text{phs} \gets [0,\, 2\pi\cdot\text{rand}( 1,\,\text{len}(S_\text{amp})-2 ),\, 0]$
		\State $X \gets S_\text{amp}\cdot\exp(S_\text{phs}) $
		\State $X \gets \left[ X,\, \overline{X(\text{end}:2)} \right]$
		\State $y = \text{ifft}(X)$
		\State $\phi_\text{vib} = y\cdot\text{len}(X)/ 2$
		\EndFunction
	\end{algorithmic}\label{alg:iFFT}
\end{algorithm}

\section{Turbulence estimation} \label{sec:turbulence}
Figure \ref{fig:Turb_est} shows the estimation of turbulence states.
For the reference signal, an open-loop simulation in {COMPASS} isolates atmospheric turbulence by setting the \ac{DM} commands and the additionally added vibrations to zero.
The \ac{NGS} Kalman filter closely tracks the reference turbulence with a delay, whereas the \ac{LGS} Kalman filter follows significantly less accurately.
The multi-rate observer adopts the trend of the \ac{NGS} \ac{KF} while also capturing the high-frequency components of the \ac{LGS}.
%
%
\begin{figure}
	\centering
	\input{figures/plots/Multi_turb1_tip.tikz}
	\\ \vspace{0.75em}
	\input{figures/plots/Multi_turb1_tilt.tikz}

	\caption[Estimation of the atmospheric turbulence.]{Estimation of the atmospheric turbulence.
		Estimation of the \ac{LGS} \ac{KF} (\protect\tikz[baseline=-0.5ex]\protect\draw[LGS, thick] (0,0) -- (0.5,0);),
		estimation of the \ac{NGS} \ac{KF} (\protect\tikz[baseline=-0.5ex]\protect\draw[NGS, thick] (0,0) -- (0.5,0);),
		estimation of the multi-rate \ac{KF} (\protect\tikz[baseline=-0.5ex]\protect\draw[Mult, thick] (0,0) -- (0.5,0);)
		and reference (\protect\tikz[baseline=-0.5ex]\protect\draw[Ref, thick, dash pattern=on 2pt off 2pt] (0,0) -- (0.5,0);).}
	\label{fig:Turb_est}
\end{figure}

\subsubsection*{Discussion of the results}
The simulation results show that the \ac{NGS} Kalman filter provides a more accurate estimation of turbulence states compared to the \ac{LGS} Kalman filter.
The limitations of the \ac{LGS} simulation, including the cone effect and the simplistic averaging method used to combine the \ac{LGS} data, cause the less accurate turbulence estimation. 
Accurate reconstruction would require a higher number of \acp{LGS} and more complex fusion of the \ac{LGS} \ac{WFS} signals, considering the geometry and shape of the light cones.
Moreover, in real situations, global tip-tilt turbulence cannot be measured by \acp{LGS} due to issues like uplink-turbulence \cite{AndrewReeves2016}.
Therefore, it is advisable to rely on turbulence estimation through \ac{NGS} \ac{WFS} data for tip-tilt modes.
Furthermore, the atmospheric \mbox{tip-tilt} variations are predominantly low-frequency, negating the need for \ac{LGS} \cite{Hardy1998}.
Conversely, higher-order modes exhibit more rapid atmospheric fluctuations, which are often not measurable by the slow \ac{NGS} \ac{WFS}.
The multi-rate observer demonstrates potential by integrating the strengths of both \ac{NGS} and \ac{LGS} measurements, but the low quality of the \ac{LGS}-based estimation severely limits its potential.
However, the multi-rate approach can be advantageous for higher-order turbulence compensation in \ac{AO} systems, where \ac{LGS} measurements of the atmosphere are more accurate.

In real life modern \ac{MCAO} systems, multiple \acp{LGS} and \acp{DM} are used to mitigate the cone effect and other \ac{LGS} issues resulting in a more accurate turbulence estimation based on \ac{LGS} data.
The \ac{NGS} still provides a more accurate depiction of the considered atmospheric window, but the slow sampling rate of the corresponding \ac{WFS} does not allow the measurement of high-frequency disturbances in higher Zernike modes.
The results of this section have shown that designing a multi-rate observer based on \ac{WFS} data is feasible and can achieve the goal of following the \ac{NGS} trend while also incorporating high-frequency disturbances measured by the \ac{LGS}.
When the \ac{LGS} and \ac{NGS} estimates are close to each other (e.g. at around \qty{1}{\s} in Figure~\ref{fig:Turb_est}), the additional jitter and noise in the multi-rate estimation are significantly reduced, enhancing overall performance.
The results of this study indicate that in our simulations, the multi-rate observer does not demonstrate superior performance to the \ac{NGS}-only Kalman filter for tip-tilt reconstruction.
However, it holds potential for higher-order real-world systems where the compensation of high-frequency disturbances is critical \cite{Rosensteiner2013}.

\begin{figure}[b]
	\centering
	\input{figures/plots/GPR_vib1.tikz}
	\\ \vspace{0.5em}
	\input{figures/plots/GPR_vib2_PSD.tikz}

	\caption[Estimation of structural vibrations.]{Estimation of structural vibrations (only tip mode shown).
		Estimation of the \ac{LGS} \ac{KF} (\protect\tikz[baseline=-0.5ex]\protect\draw[LGS, thick] (0,0) -- (0.5,0);),
		estimation of the offline \ac{GPR} (\protect\tikz[baseline=-0.5ex]\protect\draw[GPRoffl, thick] (0,0) -- (0.5,0);),
		estimation of the online \ac{GPR} (\protect\tikz[baseline=-0.5ex]\protect\draw[GPRonl, thick] (0,0) -- (0.5,0);)
		and real vibration (\protect\tikz[baseline=-0.5ex]\protect\draw[Ref, thick, dash pattern=on 2pt off 2pt] (0,0) -- (0.5,0);).}
	\label{fig:Vib_est}
\end{figure}

\section{Vibration estimation} \label{sec:vibrations}
Similar to the turbulence estimation, we focus on the tip-tilt modes, as they have the biggest impact on image quality.
The proposed multi-rate observer will only be used to estimate atmospheric turbulence, as we presume that the \ac{NGS} provides a more precise representation of the atmospheric conditions.
Both \ac{NGS} and \ac{LGS} provide equivalent information regarding structural vibrations, and the inclusion of the slower \ac{NGS} \ac{WFS} data is not expected to improve estimation performance. 
Consequently, the simulation results for the offline and online \ac{GPR} sensing vibrations are illustrated in Figure~\ref{fig:Vib_est}.
Evaluating the state~$\hat{x}_3$ of the Kalman filter using \ac{LGS} \ac{WFS} data, we get a baseline estimation for the vibration acting on the \ac{AO} loop.
However, the model inside \ac{KF} only features a single natural frequency (c.f. Table~\ref{tab:Vib_Parameter}).
Hence, the corresponding vibration estimation is often out of phase and overshoots the reference regularly.
The incorporation of more natural frequencies into the \ac{KF} model is likely to result in better estimation performance.
We calculate \iac{RMS} error of \num{16.6e-3} for the \ac{LGS} \ac{KF}.
The offline \ac{GPR} is much more accurate ($\text{\ac{RMS} error}=\num{12.4e-3}$) and closer to the real vibration \ac{PSD} for frequencies between \num{30} and \qty{65}{\Hz}.
The vibration estimation by the online \ac{GPR} has a little lower amplitude and a slightly higher \ac{RMS} of \num{13.3e-3}.
Upon closer examination, it becomes evident that the offline realization exhibits a smoother curve.

Notice that the \ac{GPR} has many parameters to tune (see Table~\ref{tab:gpr_params}) and the quality of the estimation is greatly affected by the choice of values.
It is conceivable, that for a different set of parameters, the online \ac{GPR} may show superior performance to the offline implementation.
For this particular setting, we conclude that the offline \ac{GPR} is preferable due to a slightly better \ac{RMS} and a faster computation time.

\section{Conclusion}\label{sec:conclusion}
We have introduced an \ac{AO} system description that uses linear \ac{AR2} models to describe turbulence and vibrations.
The \ac{DM} and \acp{WFS} are modeled as delay of one time step, resulting in a linear, discrete state space system containing six states.
A steady-state Kalman filter, similar to those used in real AO systems, serves as a baseline for the subsequent analyses.
Moreover, two novel disturbance observer designs have been presented by the authors.
The multi-rate Kalman filter integrates measurements from multiple different \acp{WFS} aiming to combine advantages of \ac{NGS}- and \ac{LGS}-based estimations.
For sensing structural vibrations, we have employed two variations of a model-free Gaussian process estimator together with a Butterworth filter, which is parameterized using real \ac{VLT} \ac{NGS} data.
The offline \ac{GPR} uses open-loop training data to learn the correlation between the inputs and disturbances.
In contrast, the online implementation leverages past measurement data and calculates the necessary matrices at each time step.
Both observers have been validated with data generated by the \ac{AO} simulation tool \ac{COMPASS}.

The findings of this work demonstrate the capability of the proposed methods to sense disturbances for ground-based telescopes.
The presented multi-rate filter maintains the advantages of an \ac{NGS}-only scheme while operating at the \ac{LGS} \ac{WFS} bandwidth.
Furthermore, the \ac{GPR}-based techniques offer better performance compared to state-of-the-art Kalman-based filters.
It is imperative that the observers are now integrated into the \ac{AO} loop to evaluate their optical performance gain in closed-loop.
To conclude, both approaches have yielded promising results and are ready to be extended to other scenarios and validated on a real test bench.

\acknowledgments 
The authors would like to thank the German Federal Ministry of Education and Research (BMBF) for supporting this work under grant 05A23VS1 as well as the colleagues at Institute for System Dynamics (University of Stuttgart) and Max Planck Institute for Astronomy for their helpful suggestions and comments.

\bibliography{bibliography_PJ} 
\bibliographystyle{spiebib} 

\end{document}